\documentclass[aps,prb,twocolumn,superscriptaddress]{revtex4-2}
\usepackage{graphicx}
\usepackage{amsmath}
\usepackage{amssymb}
\usepackage{mathtools}
\usepackage{xcolor}
\usepackage{braket}
\usepackage{soul}
\usepackage[colorlinks=true, citecolor=blue, urlcolor=blue, linkcolor=blue,bookmarks=false,hypertexnames=true]{hyperref} 
\usepackage{appendix}
\usepackage{array}
\usepackage[column=O]{cellspace}
\newcolumntype{P}[1]{>{\centering\arraybackslash}p{#1}}

\begin{document}
\graphicspath{}
\preprint{APS/123-QED}

\title{Optical Properties of Gated Bilayer Graphene Quantum Dots with Trigonal Warping}
\date{\today}

\author{Matthew Albert}
\thanks{malbe058@uottawa.ca}
\affiliation{Department of Physics, University of Ottawa,
Ottawa, K1N6N5, Canada}

\author{Daniel Miravet}
\thanks{dmiravet@uottawa.ca}
\affiliation{Department of Physics, University of Ottawa,
Ottawa, K1N6N5, Canada}

\author{Yasser Saleem}
\affiliation{Institut f\"{u}r Physikalische Chemie, Universit\"{a}t Hamburg, Grindelallee 117, D-20146 Hamburg, Germany}

\author{Katarzyna Sadecka}
\affiliation{Department of Physics, University of Ottawa,
Ottawa, K1N6N5, Canada}
\affiliation{Institute of Theoretical Physics, Wroc\l aw University of Science and Technology, Wybrze\.ze Wyspia\'nskiego 27, 50-370 Wroc\l aw, Poland}

\author{Marek Korkusinski}
\affiliation{Department of Physics, University of Ottawa,
Ottawa, K1N6N5, Canada}
\affiliation{Security and Disruptive Technologies,
National Research Council, Ottawa, K1A0R6, Canada }

\author{Gabriel Bester}
\affiliation{Institut f\"{u}r Physikalische Chemie, Universit\"{a}t Hamburg, Grindelallee 117, D-20146 Hamburg, Germany}

\author{Pawel Hawrylak}
\affiliation{Department of Physics, University of Ottawa,
Ottawa, K1N6N5, Canada}

%%%%%%%%%%%%%%%%%%%%%%%%%%%%%%%%%%%%%%%%% Abstract %%%%%%%%%%%%%%%%%%%%%%%%%%%%%%%%%%%%%%%

\begin{abstract}
We determine the optical properties of gated bilayer graphene quantum dots with trigonal warping (TW) of single-particle energy spectra. The lateral structure of metallic gates confines electrons and holes in a quantum dot (QD) electrostatically.
The gated bilayer graphene energy spectrum is characterized by two K-valleys surrounded by three minivalleys with energies depending on the applied vertical electric field. Employing an atomistic tight-binding model, we compute the single-particle QD states and analyze the influence of TW on the energy spectrum as the lateral confining potential depth varies. We find a regime where the QD levels are dominated by the presence of three minivalleys around each K-valley. Next, we compute dipole matrix elements and analyze the oscillator strengths and optical selection rules for optical valence to conduction band transitions. We then include electron-electron interactions by first computing the microscopic Coulomb matrix elements, electron self-energy, and solving the Bethe-Salpeter equation to obtain the excitonic spectrum. Finally, we obtain the absorption spectrum for a shallow confining potential depth, which further amplifies the effects of TW on the optical properties. Our results predict the existence of two degenerate bright exciton states, each built of the three minivalley states that do not exist in the deep confinement regime, where the effects of TW are negligible.

\vspace{6mm}

%Corresponding author email: malbe058@uottawa.ca
\end{abstract}
\maketitle

%%%%%%%%%%%%%%%%%%%%%%%%%%%%%%%%%%%% Introduction %%%%%%%%%%%%%%%%%%%%%%%%%%%%%%%%%%%%%%%%

%%%%%%%%%%%%%%%%%%%%%%%%%%%%%%%%%%%%%%%%%%%%%%%%%%%%%%%%%%
\section{Introduction}
%%%%%%%%%%%%%%%%%%%%%%%%%%%%%%%%%%%%%%%%%%%%%%%%%%%%%%%%%%
While bilayer graphene (BLG) is gapless, applying a perpendicular electric field opens the gap, making BLG a voltage tunable semiconductor \cite{castro2007biased,zhang2009direct,gava2009ab,ohta2006controlling,mak2009observation,PhysRevB.74.161403,wang2008gate,PhysRevB.75.155115}. Optical properties of semiconductors are controlled by excitons. The tunability of the bandgap with the gate motivated the study of excitons in biased BLG \cite{ju2017tunable,park2010tunable,henriques2022absorption,sauer2022exciton}. An exciton is a quasi-particle, bound state of interacting electron and a hole that forms upon optical excitation of an electron to the conduction band, leaving behind a hole in the valence band, which then interacts via Coulomb attraction. Excitons play an important role in the optical response of semiconductors, and those coupled strongly to light are particularly of interest \cite{thureja2022electrically,henriques2022absorption}. Excitons in nanoscale semiconductors, e.g., self-assembled and graphene quantum dots (QDs), have been studied due to their potential applications, including lasers, light emitting diodes (LEDs), storage devices, and sources of entangled photon pairs via biexciton-exciton cascade \cite{ozfidan2014microscopic,bayer2000hidden,PhysRevLett.85.389,sun2015biexciton,ozfidan2015theory,PhysRevB.87.115310,PhysRevB.67.161306,lundstrom1999exciton}. However, the optical properties of these QDs are controlled by confining both electrons and holes through their structural attributes: size, shape, edge type and the material composition, which is challenging to tune once fabricated. 
On the other hand, highly tunable laterally gated QDs confine either electrons or holes. Surprisingly, lateral gates can be applied to BLG in a way that leads to the emergence of a QD that effectively confines both electrons and holes \cite{pereira2007tunable,korkusinski2023spontaneous,saleem2023theory,sadecka2023electrically}. BLG QDs offer high tunability of the optical and electronic properties through lateral gate voltages and provide an ideal system for studying electrically tunable excitons. 

Recently, the effects of trigonal warping (TW) in biased Bernal-stacked BLG have become of interest \cite{seiler2024probing,zhou2022isospin,de2022cascade,dong2023isospin,seiler2022quantum,PhysRevB.101.161103}. The interplay of TW with strong electron-electron interactions, van Hove singularities, and a large applied electric field has led to isospin phase transitions \cite{zhou2022isospin,de2022cascade,dong2023isospin}, along with correlated phases exhibiting metallic and insulating properties \cite{seiler2022quantum}. In gated BLG QDs, which exhibit $C_3$ symmetry due to TW, it was found that under specific conditions, three-fold degenerate QD energy levels in each K-valley can arise \cite{knothe2020quartet, garreis2021shell}. This three-fold degeneracy results from the presence of the three minivalleys around each K-valley. While its effects have been studied in the context of transport measurements, resulting in the bunching of twelve conductance resonances \cite{garreis2021shell}, its impact on excitonic and optical properties remains unexplored.

In this work, we extend our prior theory of excitons in a gated BLG QD \cite{saleem2023theory} to include TW and its effect on optical properties. 
The paper is organized as follows. We begin by employing an atomistic tight-binding model, as outlined in Sec.~\ref{section:Model}, to describe the electronic states of biased BLG. Next, we impose a confining potential and obtain the QD energy spectrum. Subsequently, in Sec.~\ref{section:DME}, we explore the coupling of different QD states with light through dipole matrix elements (DMEs) and optical selection rules. In Sec.~\ref{section:exciton}, we discuss the effects of electron-electron interactions: Coulomb matrix elements, self-energy, and solution of the Bethe-Salpeter equation.  We discuss the combined effect of confining potential depth and the presence of minivalleys on the resulting QD level degeneracies due to TW  on the excitonic spectrum. Finally, we present the excitonic absorption spectra in Sec.~\ref{section:Absorption} and identify the optical signatures of strong and weak TW.   
%%%%%%%%%%%%%%%%%%%%%%%%%%%%%%%%%%%%%%%%% Summary %%%%%%%%%%%%%%%%%%%%%%%%%%%%%%%%%%%%%%%%
\section{Model}\label{section:Model}
\subsection{Gated Bilayer Graphene}
\begin{figure}[t]
    \centering  \includegraphics[width=\linewidth]{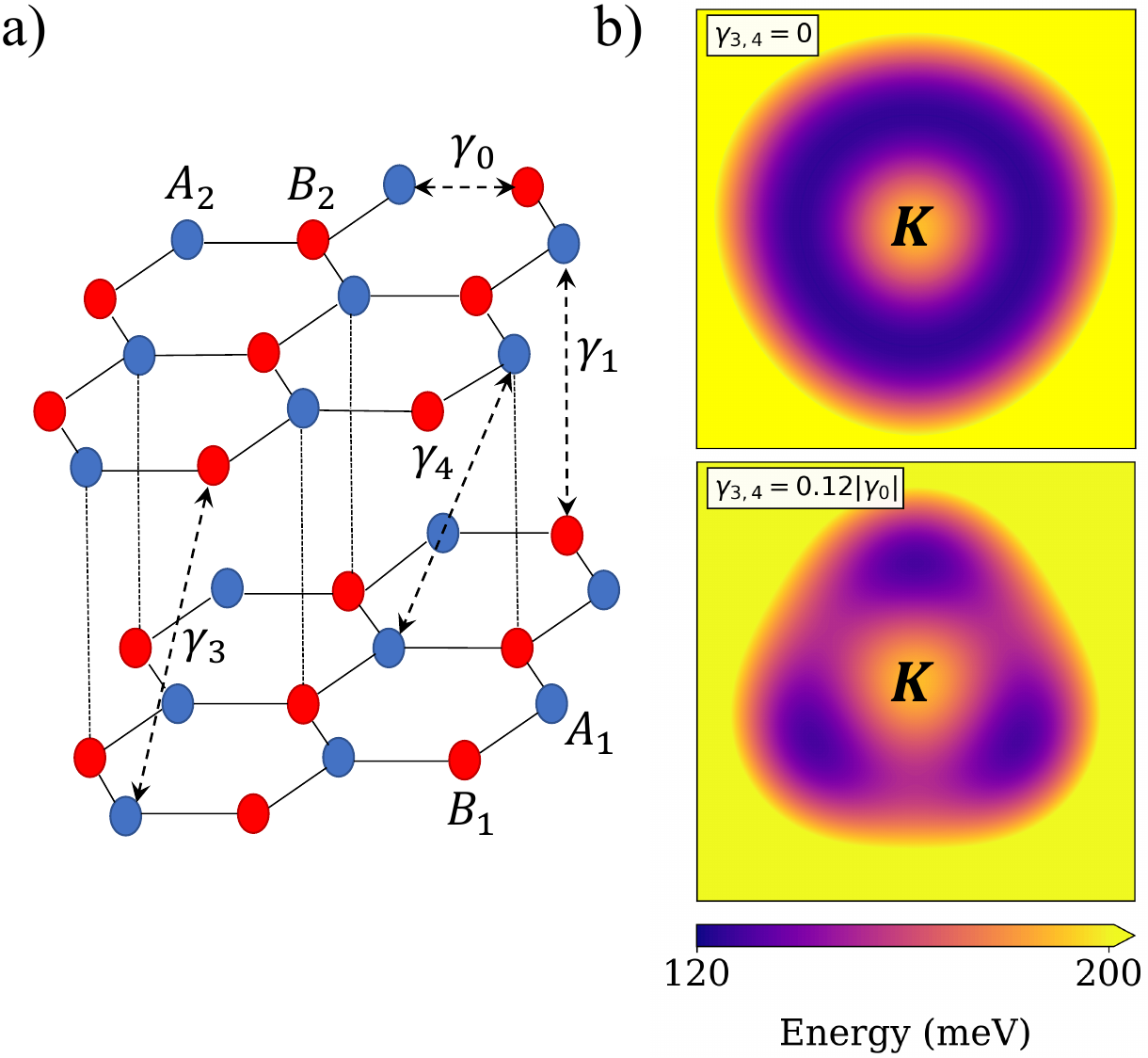}
    \caption{Band structure of gated AB-stacked BLG. (a) 3-dimensional view of AB-stacked BLG with atoms from sublattices $A_1$ and $B_1$ ($A_2$ and $B_2$) on the bottom (top) layer. (b) the lowest CB of the bulk band structure of BLG in the vicinity of K point without (top) and with (bottom) the inclusion of the TW  and of a displacement voltage of $V_E = 380$ meV.}
    \label{fig:BandStructure}
\end{figure}

Building upon our previous works \cite{saleem2023theory,korkusinski2023spontaneous,sadecka2023electrically}, we consider Bernal-stacked BLG as illustrated in Fig.~\ref{fig:BandStructure}(a), where the bottom (top) layer consists of sublattices $A_1$ and $B_1$ ($A_2$ and $B_2$). The nearest neighbor (NN) in-plane bond length is $a=0.143$~nm, and the interlayer distance is $h=0.335$~nm. Our selected unit cell vectors are defined as $\Vec{a}_1=a(0,\sqrt{3})$ and $\Vec{a}_2=\frac{a}{2}(3,-\sqrt{3}).$ We model our BLG graphene by a rhomboidal computational box constructed from a maximum of $N_1=N_2=1133$ unit cells along the directions $\Vec{a}_1$ and $\Vec{a}_2$, encompassing over 5.1 million carbon atoms. Next, periodic boundary conditions are imposed on the rhombus to eliminate finite-size effects, resulting in a discrete set of $k$-points in momentum space \cite{sadecka2023electrically}. Utilizing Bloch's theorem, the wavefunction on a sublattice $l$ is expressed as $\ket{\phi^l_{\Vec{k}}}= \frac{1}{\sqrt{N}}\sum_{\Vec{R}_l} e^{i\Vec{k}\cdot \Vec{R_l}}\ket{\Vec{R}_l}$, where $N=N_1\times N_2$ is the total number of unit cells and $\ket{\Vec{R}_l}$ represents a $p_z$ orbital localized on an atom at position $\Vec{R}_l.$ To open the energy gap, we apply an external electric field perpendicular to the layers, resulting in a potential of $V_E/2$ ($-V_E/2$) on the bottom (top) layer and an overall potential difference between layers of $V_E$. The Hamiltonian describing bulk BLG in the presence of the applied electric field expressed in the basis of sublattices is given by \cite{mccann2013electronic,PhysRevLett.96.086805,PhysRevB.77.195403,PhysRevB.74.075404}:
\begin{equation}\label{eq:BulkHamiltonian}
H_{\textrm{bulk}}(\Vec{k}) = 
    \begin{pmatrix}
        \frac{V_E}{2} & \gamma_0 f(\Vec{k}) & \gamma_4 f(\Vec{k}) & \gamma_3 f^*(\Vec{k})\\
         \gamma_0 f^*(\Vec{k})  & \frac{V_E}{2} & \gamma_1 &
        \gamma_4 f(\Vec{k}) \\ \gamma_4 f^*(\Vec{k}) & \gamma_1 &  -\frac{V_E}{2} &  \gamma_0 f(\Vec{k}) \\
        \gamma_3 f(\Vec{k}) & \gamma_4 f^*(\Vec{k}) & \gamma_0 f^*(\Vec{k}) & -\frac{V_E}{2} \\
    \end{pmatrix},
\end{equation}
where the parameter $\gamma_0=-2.5$~eV represents the NN intralayer hopping and $\gamma_1=0.34$~eV denotes the interlayer hopping between pairs of orbitals on sublattices $B_1$ and $A_2$  \cite{sadecka2023electrically,saleem2023theory}. The parameters $\gamma_3$ and $\gamma_4$ describe the next NN interlayer hopping but differ in that $\gamma_3$ represents the next NN interlayer hopping between pairs of orbitals localized on unstacked atoms, whereas $\gamma_4$ is between an orbital on a stacked and an unstacked atom, as illustrated in Fig.~\ref{fig:BandStructure}(a). The inclusion of $\gamma_3$ introduces the effects of TW, whereas $\gamma_4$ breaks the electron-hole symmetry \cite{li2009band,zhang2008determination}. We set $\gamma_3 = \gamma_4 = 0.12|\gamma_0|$ \cite{korkusinski2023spontaneous,sadecka2023electrically}. The function $f(\Vec{k})$ is given as $f(\Vec{k})=\sum_{j=1}^3 e^{i\Vec{k}\cdot \Vec{\delta_j}}$, where  $\Vec{\delta}_j$ are the positions of the three NNs of a given atom. We can write them as $\Vec{\delta}_1 = \Vec{b}, \, \Vec{\delta}_2 = \Vec{b}-\Vec{a}_2-\Vec{a}_1$ and $\Vec{\delta}_3 = \Vec{b}-\Vec{a}_1$ with $\Vec{b} = \frac{a}{2}(1,\sqrt{3}).$  For a particular $\vec{k}$, we diagonalize the $4\times4$ Hamiltonian in Eq.~(\ref{eq:BulkHamiltonian}) obtaining band eigenstates of the form $\ket{\psi^p_{\Vec{k}}}=\sum_{l} A^p_{\Vec{k},l} \ket{\phi_{\Vec{k}}^l},$ where $p$ labels one of the four bands (two valence bands (VBs) and two conduction bands (CBs)), and $l$ represents the sublattices $A_1, B_1, A_2$ and $B_2.$ The upper panel of Fig.~\ref{fig:BandStructure}(b) displays a heatmap of the lowest CB energy near the K point, with $\gamma_3 = \gamma_4 = 0$ and $V_E = 380$ meV. In the absence of TW, a ring of low-energy minima emerges, which exhibits rotational symmetry and resembles the shape of a Mexican hat \cite{knothe2018influence,varlet2015tunable}. Upon incorporating TW effects, the symmetry of our system reduces to $C_3$, resulting in the three minivalleys forming around the K point, as depicted in the lower plot of Fig.~\ref{fig:BandStructure}(b). Increasing the strength of the displacement field enlarges the band gap, causing the three minivalleys to shift away from each other and accentuating the depth of the three minivalleys surrounding the K and K' points \cite{garreis2021shell,seiler2022quantum}. 

%%%%%%%%%%%%%%%%%%%%%%%%%%%%%%%%%%%%%%%%%%%
\subsection{Gated Quantum Dot} 
%%%%%%%%%%%%%%%%%%%%%%%%%%%%%%%%%%%%%%%%%%%

\begin{figure*}[t]
    \centering  \includegraphics[width=1.0\textwidth]{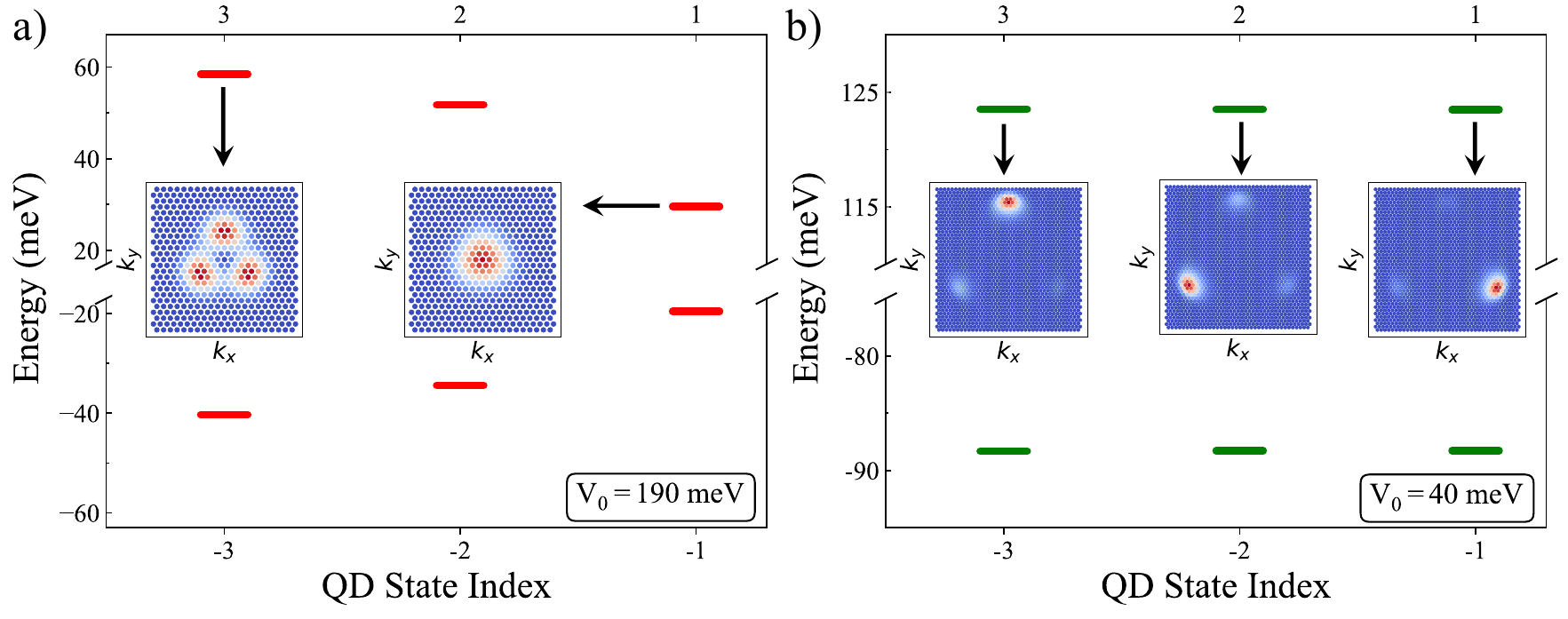}
    \caption{Single-particle energy spectrum of gated bilayer graphene quantum dot near the Fermi level with a QD radius of $R_{QD} = 20 $ nm, displacement voltage of $V_E=380$ meV and confining potential depth of (a) $V_0 = 190$ meV and (b) $V_0=40$ meV. Insets present the Fourier transform of the charge density in momentum space for QD states in valley K.}
    \label{fig:QD Spectrum}
\end{figure*}

We apply different lateral confining potentials to each layer, effectively confining holes on the top layer ($z=+\frac{h}{2}$) and electrons on the bottom layer ($z=-\frac{h}{2}$) \cite{sadecka2023electrically,saleem2023theory,korkusinski2023spontaneous}. To maintain simplicity, we adopt a single Gaussian potential given by: 
\begin{equation}\label{eq:ConfiningPotential}
    {V}_{QD}(\rho) =
    \begin{cases}
        +V_0 \, e^{-\frac{\rho^2}{R_{QD}^2}}, \quad z = +\frac{h}{2}, \\
        -V_0 \, e^{-\frac{\rho^2}{R_{QD}^2}}, \quad z = -\frac{h}{2},
    \end{cases}
\end{equation}
where $R_{QD}$ is the QD radius and $V_0$ represents the depth of the confining potential. The confining potential given in Eq.~(\ref{eq:ConfiningPotential}) resembles a parabolic confining potential \cite{que1992excitons,zarenia2013electron,pereira2007tunable,kyriakidis2002voltage} near the center of the dot, but differs in that it converges to zero at large distances from the center of the dot. The Hamiltonian for the gated BLG QD becomes:
\begin{equation}\label{eq:QDHamiltonian}
    \hat{H}_{QD} = \hat{H}_{\textrm{bulk}} + \hat{V}_{QD}.
\end{equation}
To obtain the single-particle eigenstates of the QD Hamiltonian, we expand the QD states in the basis of band eigenstates of the bulk Hamiltonian in Eq.~(\ref{eq:BulkHamiltonian}), $\varphi_s(\Vec{r})= \sum_{\Vec{k}}\sum_{p} B^s_{p,\Vec{k}} \, \psi^p_{\Vec{k}}(\Vec{r})$, where $\Vec{k}$ represents the discrete set of points in the momentum space and $p$ denotes the BLG band index. Acting with the QD Hamiltonian in Eq.~(\ref{eq:QDHamiltonian}) on the single-particle QD states gives an eigenvalue equation:
\begin{equation}\label{eq:Schrodinger}
    \varepsilon_{p,\Vec{k}}\, B^s_{p,\Vec{k}} +\sum_{p', \Vec{k}'}\braket{\psi_{\Vec{k}}^p|\hat{V}_{QD}|\psi_{\Vec{k}'}^{p'}}B^s_{p',\Vec{k}'}=E_s B^s_{p,\Vec{k}},
\end{equation}
where $\varepsilon_{p,\Vec{k}}$ are the eigenenergies of the bulk Hamiltonian and the coefficients $B^s_{p,\Vec{k}}$ are obtained by solving Eq.~(\ref{eq:Schrodinger}). The confining potential couples wavevectors $\Vec{k}$ and $\Vec{k}'$ on bands $p$ and $p'$ via the matrix element:
\begin{equation}
\braket{\psi_{\Vec{k}}^p|\hat{V}_{QD}|\psi_{\Vec{k}'}^{p'}} = \sum_{l} (A^{p}_{\Vec{k},l})^*\, A^{p'}_{\Vec{k}',l} e^{i\left(\Vec{k}'-\Vec{k}\right)\cdot \Vec{d}_l} V_{\Vec{k},\Vec{k}',l},
\end{equation}
where $\Vec{d}_l$ is the relative position of an atom on sublattice $l$ within a unit cell, and $V_{\Vec{k},\Vec{k}',l}$ is the Fourier transform of the confining potential as defined in Eq.~(\ref{eq:ConfiningPotential}) specifically for sublattice $l$. The explicit expression for $V_{\Vec{k},\Vec{k}',l}$ is given by:
\begin{equation}\label{eq:FourierTransform}
    V_{\Vec{k},\Vec{k}',l} = \pm V_0 \frac{R_{QD}^2}{4\pi} A_k e^{-\frac{R_{QD}^2 |\Vec{k}-\Vec{k}'|^2}{4}},
\end{equation}
where $A_k = \frac{8\pi^2}{3\sqrt{3}a^2N}$ represents the reciprocal lattice unit cell area, and the $+$ ($-$) sign corresponds to sublattices $l=A_2, B_2$, $(l=A_1, B_1)$. We impose an energy cut-off of $E_{cut}=600$ meV on our band states $\ket{\psi_{\Vec{k}}^p}$, including only those band states with energy $\varepsilon_{p,\Vec{k}}$ in the range $|\varepsilon_{p,\Vec{k}}| \leq E_{cut}$. This ensures the convergence of single-particle QD energies and wavefunctions built with $\vec{k}$-vectors around the Fermi level. Half of our band states are near the K point, while the other half cluster around the K' point, giving rise to valley doublets and quadruply degenerate electronic shells, including spin. 

To obtain the QD energy spectrum, we solve the eigenvalue problem defined by Eq.~(\ref{eq:Schrodinger}). The resulting energy levels for a displacement voltage of $V_E = 380$ meV, QD radius of $R_{QD}=20$~nm, and confining potential depth of $V_0 = 190$ meV, are displayed in Fig.~\ref{fig:QD Spectrum}(a). These levels are nondegenerate when neglecting spin and valley degeneracies. In the insets of Fig.~\ref{fig:QD Spectrum}(a), we present the Fourier transform of the charge density around valley K for the first and third QD CB states. The lowest energy CB states in biased BLG originate from the three minivalleys, so one might expect a similar localization for the low-energy QD CB states. Interestingly, upon examining the Fourier transform of the charge density for the lowest QD CB state, one observes that it does not localize in the three minivalleys; instead, it appears centered around the K point. The $C_3$ symmetry induced by TW is noticeable when examining higher energy levels. However, for the third QD CB state, the three maxima in the charge density do not coincide with the position of the minivalleys, which suggests that TW effects do not strongly influence the low-energy QD states.
 
Tuning BLG and QD parameters enhances TW's effect, resulting in triple degenerate QD energy levels. The complex interplay between the inherent properties of gated BLG ($C_3$ symmetry from three minivalleys) and the QD confining potential possessing $C_\infty$ symmetry governs this phenomenon. The confining potential given in Eq.~(\ref{eq:FourierTransform}) strongly couples states across momentum space for small QD radii and large confining potential depths. Notably, the three minivalleys couple, leading to low-energy QD levels localized near the center of the K point rather than the three minivalleys. As a result, the QD energy levels are singly degenerate, similar to those in parabolic confinement. Conversely, for larger QD radii and smaller confining potential depths, the confining potential weakly couples states in momentum space, leading to bulk properties manifesting more strongly in the single-particle QD energy spectrum. In this scenario, a sufficiently large applied voltage that separates the minivalleys far enough in momentum space will yield triple degenerate QD energy levels \cite{knothe2020quartet}. It is important to note that this triple degeneracy is robust and can be obtained for a large space of parameters.
 
Since the displacement field of $V_E=380$ meV is large enough to produce a significant separation of minivalleys in momentum space, we can fix $V_E$ and $R_{QD}$ while reducing the depth of our confining potential to achieve shallow confinement, setting $V_0=40$ meV. Under these conditions, a set of triple degenerate levels emerges, as depicted in Fig.~\ref{fig:QD Spectrum}(b). Upon examining the Fourier transform of the charge density for the triple-degenerate lowest CB energy levels, one observes the localization of each degenerate state around one of the three minivalleys, indicating a regime where TW's effects are significant. This reduced confinement has decreased the interlevel spacing and increased the gap between the CB and VB QD states. Additionally, this has also led to a reduction in the number of confined single-particle QD states. When selecting the confinement potential depth, one must carefully choose a depth that is shallow enough to avoid masking the effects of TW (resulting in the triple degeneracy of our QD energy spectrum) yet deep enough to yield a sufficient number of single-particle confined states, ensuring that the results are independent of the size of the computational box.

\begin{figure}[t]
    \centering  \includegraphics[width=\linewidth]{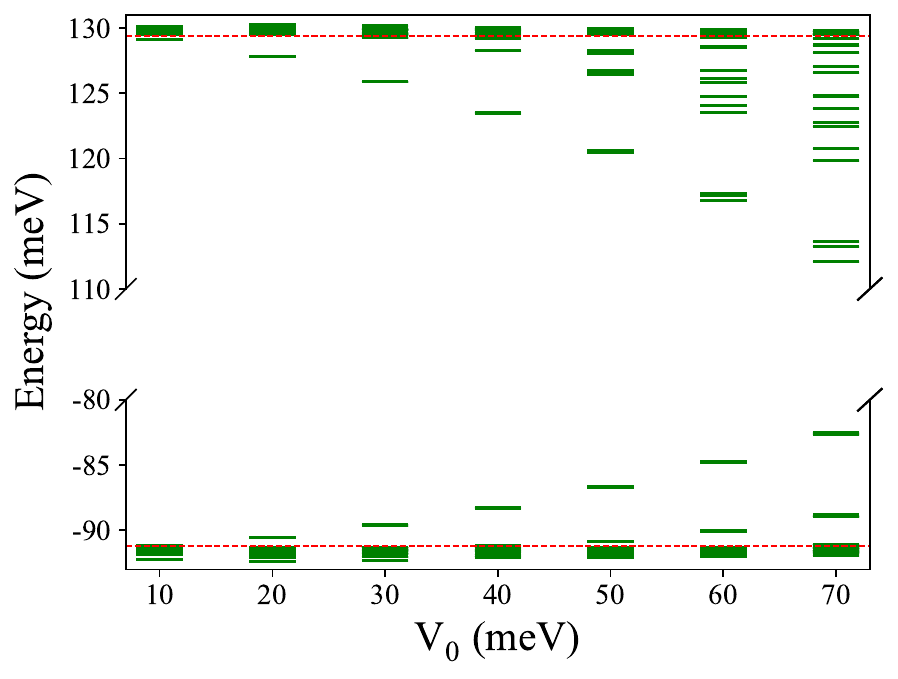}
    \caption{Single-particle energy spectrum of gated BLG QD as a function of confining potential depth within the 10 to 70 meV range, with $R_{QD}=20$~nm and $V_E=380$~meV. Green dashes represent the single-particle QD energy levels, while dashed red lines at $-91$ and $129$~meV mark the onset of the continuum in the VB and CB, respectively.}
    \label{fig:Continuum}
\end{figure}

In Fig.~\ref{fig:Continuum}, we illustrate the evolution of the single-particle QD spectrum as a function of the confining potential depth $V_0$. Green dashes represent QD energy levels, while dashed red lines mark the border between states confined in the QD and the continuum. As $V_0$ increases, QD states peel off from the continuum, becoming confined within the gap. Interestingly, the rate at which QD states peel from the continuum in the CB surpasses that in the VB. This asymmetry results from $\gamma_4$, which breaks the electron-hole symmetry. Below $V_0 = 50$ meV, the QD levels exhibit triple degeneracy. However, when $V_0=60$~meV, we observe a small splitting of the triple degenerate shells in the CB. As the confining potential depth reaches 70 meV, the previously degenerate levels in the CB separate into three distinct QD energy levels, although the triple degeneracy persists in the VB. This observation indicates that while QD states in the CB are more easily confined, maintaining the triple degeneracy is more challenging in the CB than in the VB. 

%%%%%%%%%%%%%%%%%%%%%%%%%%%%%%%%%%%%%%%%%%%%
\section{Dipole Matrix Elements} \label{section:DME}
%%%%%%%%%%%%%%%%%%%%%%%%%%%%%%%%%%%%%%%%%%

\begin{figure*}[t]
    \centering  \includegraphics[width=\textwidth]{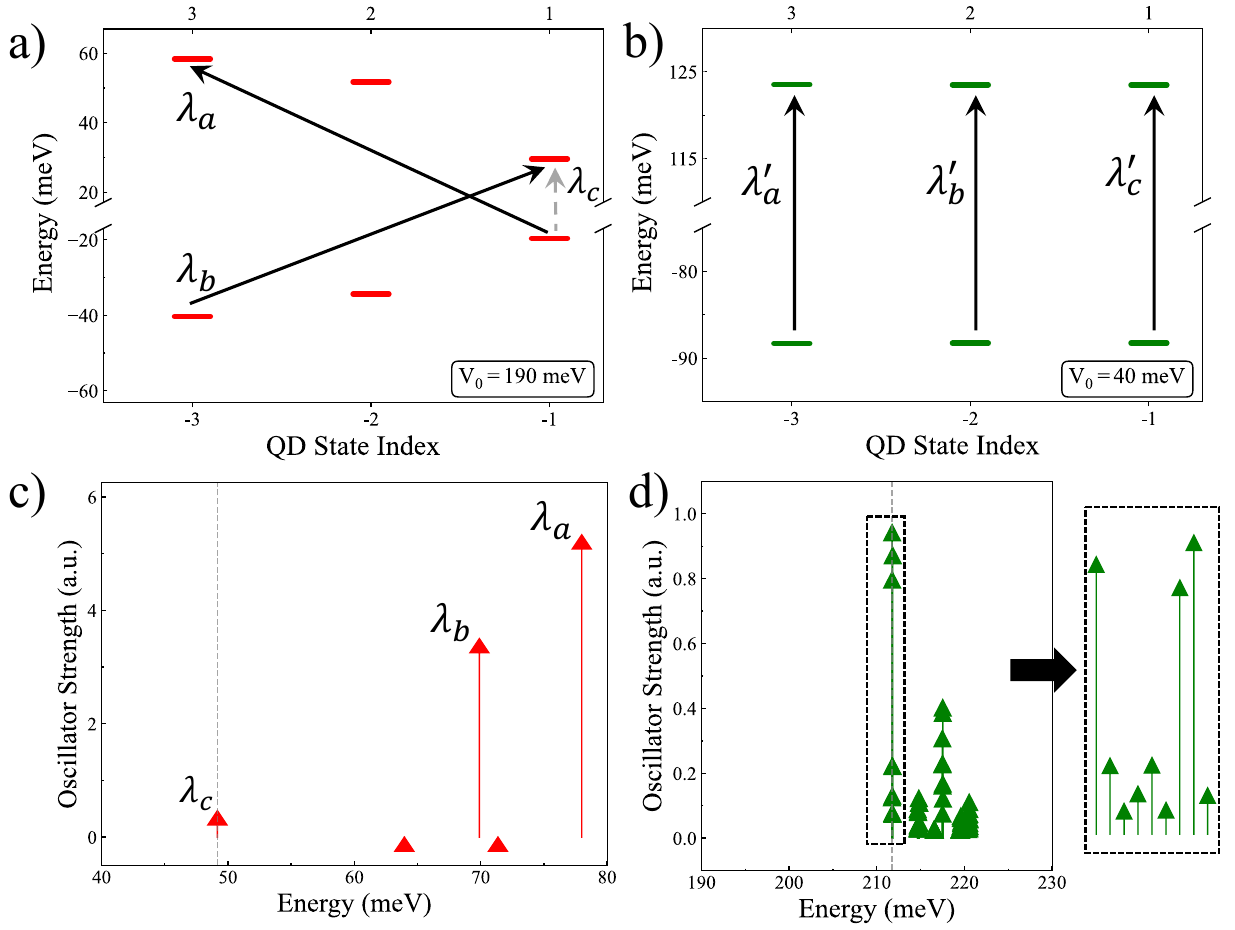}
    \caption{Dipole matrix elements for optical transitions. (a,b) Single-particle energy spectrum of gated BLG QD near the Fermi level for a single valley with $V_0=190$~meV/40 meV. Black arrows connect pairs of VB and CB states characterized by significant dipole matrix elements, and a dashed grey arrow connects the dark transition across the gap. (c,d) Joint optical density of states for the non-interacting electron and hole in BLG QD with $V_0=190$~meV/40 meV. Red/green arrows show the magnitude of the corresponding dipole elements. As the transitions occur at the same energy, we split the peaks manually to display the nine transitions to the right of (d).}
    \label{fig:DME}
\end{figure*}

Now that we have obtained the single-particle QD states, the next step is to examine the interaction of the BLG QD with light. The aim is to explore the influence of a shallow confining potential and TW on the oscillator strengths and optical selection rules. We incorporate the coupling of light with different states through dipole matrix elements $\Vec{D}_{p,q}=\braket{p| \Vec{r}|q}$, which connect the QD states $p$ and $q$, with $\Vec{r}$ denoting the electron position. In Fig.~\ref{fig:DME}(a), black arrows connect states in the VB and CB with large DMEs for a single valley, considering a deep confining potential of $V_0 = 190$~meV. In contrast to self-assembled QDs \cite{PhysRevLett.92.187402}, the state at the top of the VB (identified by the QD index $-1$) exhibits a weak connection to the lowest CB state ($+1$) but instead couples more strongly to the third excited state ($+3$). $\lambda_a$ denotes this transition. Symmetrically, the strong transition $\lambda_b$ occurs between the bottom of the CB ($+1$) and the third VB state ($-3$). One can conduct an identical analysis with QD states in the other valley. Transitions within each valley can be analyzed separately, as QD states from different valleys do not optically couple. 

In Fig.~\ref{fig:DME}(c), we present the joint optical density of states for non-interacting electron-hole pairs. We label the transitions corresponding to each peak, and as observed in our previous work \cite{sadecka2023electrically}, the brightest peaks correspond to the transitions $\lambda_a$ and $\lambda_b$. In contrast to the scenario without TW considered in Ref.~\onlinecite{saleem2023theory}, the transition across the gap $\lambda_c$ possesses a small but nonzero oscillator strength. This enhancement of the oscillator strength across the gap arises from the TW effect in the bulk Hamiltonian given by Eq.~(\ref{eq:BulkHamiltonian}), which reduces the symmetry around the K and K' points to $C_3$ \cite{ju2017tunable}. A better understanding can be derived considering a continuum model, where the difference of angular momentum $m$ between these two states involved in transition $\lambda_c$ is $+2$ $(-2)$ in valley K (K') when TW is neglected. A photon changes the total angular momentum quantum number $m$ by $\pm 1$ \cite{ju2017tunable}, which is why this transition is dark. However, with the inclusion of TW, the cylindrical symmetry around the K and K' points is broken, resulting in a difference in angular momentum between these states closer to $\pm 1$, yielding a brightening of $\lambda_c$.

After tuning the depth of the confining potential to $V_0=40$~meV, triple degenerate energy levels emerge in the VB and CB, resulting in nine possible transitions across the band gap. Three of these nine transitions are bright and labeled as $\lambda_a'$, $\lambda_b'$, and $\lambda_c'$, as illustrated in Fig.~\ref{fig:DME}(b). These three bright transitions exhibit similar oscillator strengths as displayed in Fig.~\ref{fig:DME}(d) and originate from transitions between VB and CB states localized around the same minivalley. In the subsequent sections focusing on excitons and absorption, we explore how including electron-electron interactions modifies the observed light-matter interaction and how these new selection rules impact the brightness of the low-energy exciton states.

\section{Exciton Spectrum} \label{section:exciton}
\begin{figure}[t]
    \centering  \includegraphics[width=\linewidth]{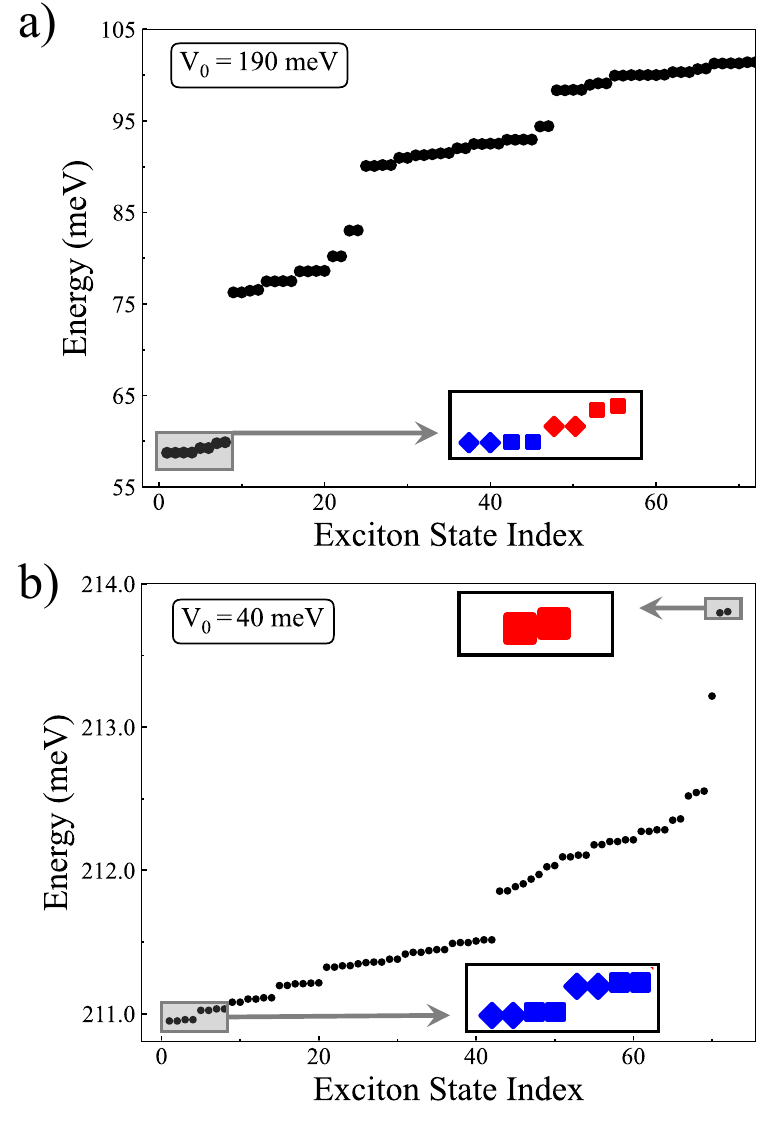}
    \caption{Excitonic energy spectra for confining potential depths of (a) $V_0=190$ meV and (b) $V_0=40$ meV. The inset boxes highlight the 8 lowest (bottom right) and 2 highest (middle right) energy states. The states are labeled as intervalley (blue), intravalley (red), singlet (square), and triplet (diamond) states. }
    \label{fig:ExcitonicEnergy}
\end{figure}

%%%%%%%%%%%%%%%%%%%%%%%%%%%%%%%%%%%%%%%%%%%%
In this section, we study excitons and analyze the effects of combined shallow confinement and TW on the exciton spectrum. We begin with the many-electron Hamiltonian written in the basis of single-particle QD states given by \cite{saleem2023theory,sadecka2023electrically}:
\begin{equation}\label{eq:ManyBody}
    \begin{aligned}
        \hat{H}_{\textrm{MB}} = & \sum_{p,\sigma}E_p c^\dagger_{p,\sigma}c_{p,\sigma} \\
        & + \frac{1}{2} \sum_{p,q,r,s}\sum_{\sigma,\sigma'} \braket{p,q|V_C|r,s} c^\dagger_{p,\sigma}c^\dagger_{q,\sigma'}c_{r,\sigma'}c_{s,\sigma} \\ &  -\sum_{p,s,\sigma} V^P_{p,s} c^\dagger_{p,\sigma}c_{s,\sigma},
    \end{aligned}
\end{equation}
where the operators $c^{\dagger}_{p,\sigma}$ $(c_{p,\sigma})$ create (annihilate) an electron in a QD state $p$ with spin $\sigma.$  The second term of Eq.~(\ref{eq:ManyBody}) accounts for the electron-electron interactions with potential $V_C = \frac{e^2}{4\pi \varepsilon_0\kappa |\Vec{r}_1-\Vec{r}_2|}$, where $\kappa$ is the dielectric constant , taken as $\kappa=6$ here \cite{gucclu2014graphene,saleem2023theory}. The Coulomb matrix elements can be written in terms of atomic orbitals and computed numerically using Slater-like $p_z$ orbitals \cite{korkusinski2023spontaneous, gucclu2014graphene}. In the last term, to maintain charge neutrality, we include a positive charge background, modeled as $V^P_{p,s}=2\sum_m^{N_{occ}}\braket{p,m|V_C|m,s}$, where $N_{occ}$ corresponds to the number of occupied states. We assume that the positive charges occupy orbitals identical to those of electrons and share the same distribution as the electrons in a filled VB. We approximate the many-electron ground state as a single Slater determinant $\ket{GS}=\prod_{p,\sigma} c^\dagger_{p,\sigma}\ket{0},$ where index $p$ runs over occupied VB states. 
The exciton wavefunction $\ket{\Psi^\mu}$ for  exciton state $ \mu$  is a linear combination of electron-hole pair excitations conserving $S_z$:
\begin{equation}
    \ket{\Psi^\mu}=\sum_{p,q,\sigma} A^\mu_{p,q,\sigma}c^\dagger_{q,\sigma}c_{p,\sigma}\ket{GS}. 
\end{equation}
Here, $\mu$ indexes the exciton states, and the index $p$ labels all VB states, while $q$ labels all CB states. The amplitudes $A^\mu_{p,q,\sigma}$ and excitonic energies $E_\mu$ can be obtained by solving the Bethe-Salpeter equation \cite{saleem2023theory}:
\begin{equation}\label{eq:BSE}
\begin{aligned}
    &\left[(E_{q}+\textstyle\sum_{q,\sigma})-(E_{p}+\sum_{p,\sigma})\right] A^\mu_{p,q,\sigma} \\ &+\sum_{p',q',\sigma'} \left[\braket{p',q|V_C|p,q'}-\braket{p',q|V_C|q',p}\delta_{\sigma,\sigma'}\right] A^\mu_{p',q',\sigma'} \\ &= E_\mu A^\mu_{p,q,\sigma}.
\end{aligned}
\end{equation}
The electron-hole pair energy is renormalized by self-energies $\Sigma_{q,\sigma}=-\sum_{p,\sigma'} \delta_{\sigma,\sigma'} \braket{p,q|V_C|p,q},$ where $p$ indexes VB states. Examining the second line of  Eq.~(\ref{eq:BSE}) with $p=p'$ and $q=q'$, one observes that the repulsive exchange interaction raises the energy of an electron-hole pair while the attractive direct interaction lowers it. We solve the Bethe-Salpeter equation in the subspace of 32 QD states around the Fermi level for a confining potential depth of $V_0 = 190$ meV. Fig.~\ref{fig:ExcitonicEnergy}(a) presents the resulting excitonic energy spectrum. 

\begin{figure*}[t]
    \centering  \includegraphics[width=\textwidth]{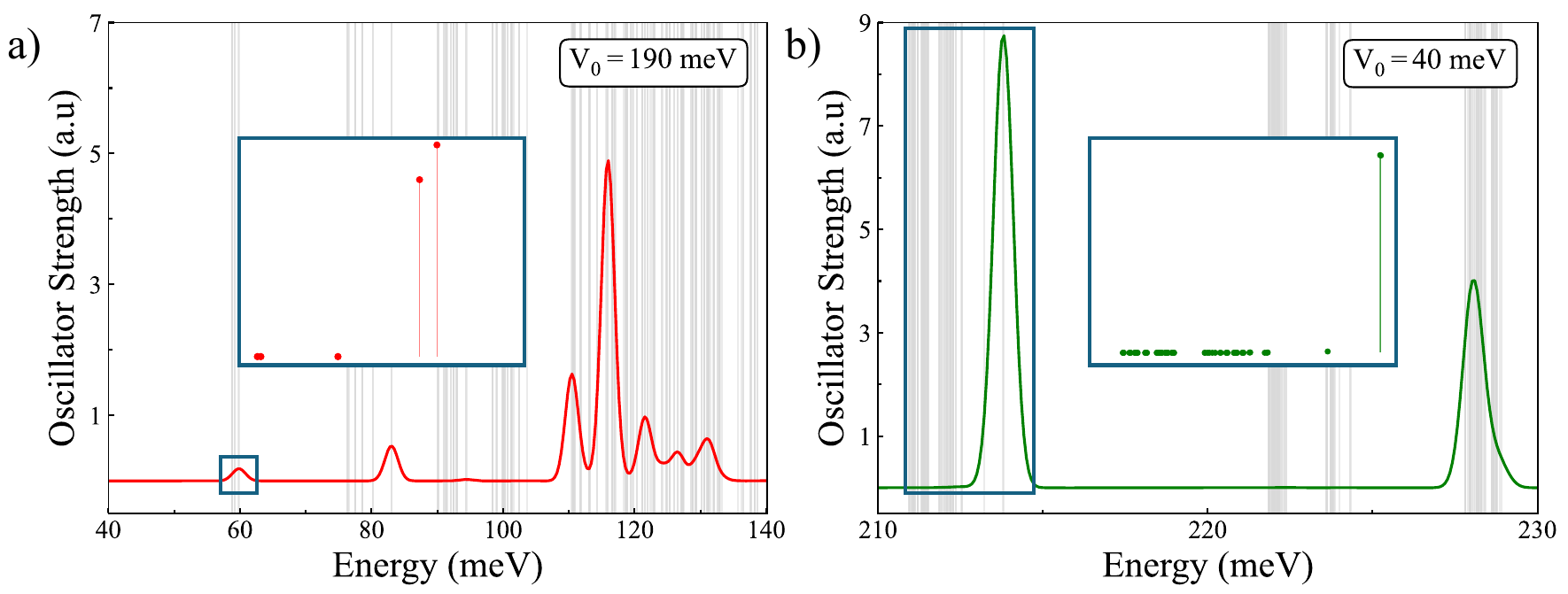}
    \caption{Exciton absorption spectrum with a confining potential depth of (a) $V_0=40$~meV and (b) $V_0=190$~meV. The solid lines depict the Gaussian broadened absorption spectrum, while the insets showcase the unbroadened spectrum, specifically focusing on the low-energy region. Grey vertical lines correspond to the energies of the excitonic states.} 
    \label{fig:Absorption}
\end{figure*}
The existence of eight low-energy exciton states can be understood by first examining noninteracting electron-hole pairs. We start by constructing electron-hole pairs from states at the top of the VB and bottom of the CB while conserving $S_z$. This involves removing an electron from the top of the VB $(-1)$ and placing this electron on the lowest CB state $(+1)$ in Fig.~\ref{fig:DME}(a). Due to the spin and valley degeneracies of the QD level, there are two possible electron-hole pairs in each valley and four pairs between valleys (including spin), resulting in eight electron-hole pair configurations constructed across the gap. These eight configurations are degenerate; however, upon the inclusion of interactions, the eight configurations mix and separate in energy. The configurations are categorized as intervalley and intravalley electron-hole pairs, where the electron and hole can either be from QD states in different or the same valleys. While the direct interaction is valley-independent, the exchange interaction is larger within a valley than between valleys. Intervalley electron-hole pairs increase the valley polarization compared to intravalley pairs, resulting in lower energy for intervalley pairs through increased exchange interaction. These excitons can be further categorized by spin as singlets or triplets, with triplets having lower energy due to exchange interaction. This is why the ground state exciton is an intervalley triplet exciton, a linear combination of intervalley electron-hole pairs with total spin $S=1$.

As previously mentioned, by tuning the depth of the confining potential to $ V_0=40~$meV, the number of confined states around the Fermi level has decreased. Upon examining Fig.~\ref{fig:Continuum}, one observes that for $V_0 = 40$, three CB and one VB states are within the gap. Since each green QD level represents three degenerate QD levels, this implies the confinement of three VB and nine CB states for a single valley. Consequently, including both valleys, we have six VB and eighteen CB confined states. Hence, we solve the Bethe-Salpeter equation in the subspace of 24 single-particle QD states around the Fermi level and present the low-energy exciton spectrum in Fig.~\ref{fig:ExcitonicEnergy}(b). These 24 QD states give the most important contributions to the low-energy exciton spectrum.

The triple degeneracy, combined with spin and valley degeneracy, leads to 72 one-pair excitations across the gap, conserving $S_z=0$. These one-pair excitations represent the dominant configurations forming the 72 low-energy exciton states. The ground state exciton remains an intervalley triplet, while the two highest energy $1s$ excitons are intravalley singlets. These two $1s$ intravalley excitons are primarily composed of one-pair excitation involving states participating in the bright transitions $\lambda_a', \lambda_b'$ and $\lambda_c'$. They are formed by one-pair excitations between QD states localized around the three minivalleys. Accordingly, we denote these two high-energy $1s$ intravalley excitons as the 'minivalley excitons'.

%%%%%%%%%%%%%%%%%%%%%%%%%%%%%%%%%%%%%%%%%%

%%%%%%%%%%%%%%%%%%%%%%%%%%%%%%%%%%%%%%%%%%
\section{Absorption Spectra} \label{section:Absorption}
%%%%%%%%%%%%%%%%%%%%%%%%%%%%%%%%%%%%%%%%%%

We now investigate the absorption of a single photon by the gated BLG QD, which is excited from the ground state to an exciton state optically. The absorption spectrum gives the probability that a photon with energy $\omega$ is absorbed  \cite{ozfidan2014microscopic,zielinski2010atomistic,ozfidan2015theory,saleem2023theory} as: 
\begin{equation}\label{eq:Absorption}
    A(\omega)= \sum_{\mu}\left|\sum_{p,q,\sigma} \Vec{\epsilon}_{\pm}\cdot \Vec{D}_{q,p}(A^\mu_{p,q,\sigma})^*\right|^2 \delta(E_\mu-\omega).
\end{equation}
Here, we consider circularly polarized light with $\Vec{\epsilon}_\pm = \frac{1}{\sqrt{2}}(1,\pm i),$ allowing for both left and right-handed polarizations. We assumed the initial state to be the ground state, corresponding to a fully occupied VB. Fig.~\ref{fig:Absorption}(a) displays the absorption spectrum obtained from Eq.~(\ref{eq:Absorption}) for a confining potential depth of $V_0=190$~meV. The grey vertical lines correspond to the excitonic energies $E_\mu$, and the red line corresponds to the Gaussian broadened absorption spectrum. Notably, a small peak around 60 meV is observed for the low-energy exciton, corresponding to the small DME of the dark transition $\lambda_c$ as observed in Fig.~\ref{fig:DME}(c). However, at higher energies, around 116 meV,  one observes a large peak corresponding to the optically bright transitions $\lambda_a$ and $\lambda_b.$

We now contrast this scenario with the absorption spectrum obtained for a QD with a shallow confining potential depth of $V_0 = 40$ meV, depicted in Fig.~\ref{fig:Absorption}(b). Here, one notices a large peak at around 214 meV, corresponding to optical transitions $\lambda_a', \lambda_b'$, and  $\lambda_c'$. An inset displays the unbroadened absorption spectrum; the bright peak arises due to the two minivalley excitons. Remarkably, these two minivalley excitons are the only bright $1s$ excitons, while the remaining seventy $1s$ exciton states are dark. This phenomenon arises from electron-electron interactions, which split exciton states into intervalley and intravalley excitons. Intervalley excitons consist of electron-hole pairs originating from different valleys connected by vanishing dipole matrix elements, rendering them dark. The excitons group into spin singlets and triplets, with triplets being dark. 

Moreover, an exciton can only be bright if the dominant electron-hole pair configurations consist of optically active states connected by DMEs. The two minivalley excitons are intravalley singlet excitons built predominantly of one-pair excitations involved in the bright transitions and, thereby, are bright. The abundance of dark, low-energy excitons highlights the potential of BLG QDs as an excellent candidate for efficient photon storage \cite{saleem2023theory}. Furthermore, the energy difference between the ground state and the first bright exciton state has decreased, potentially allowing for control over the radiative lifetime of these excitons \cite{sauer2022exciton}, thus offering customizable emission properties in the THz photon energy regime.

\section{Summary}
In summary, we presented a theory of optical properties of gated BLG QDs, including TW.  Using an atomistic tight-binding model, we computed the QD energy spectrum. We identified the emergence of triple degenerate QD energy levels due to shallow confinement and TW interplay. Subsequently, we computed dipole matrix elements and determined optical selection rules. We computed the Coulomb matrix elements and self-energies, solved the Bethe-Salpeter equation, and determined the exciton spectrum from weak to strong TW controlled by gates.  Notably, we predicted the existence of two bright $1s$ minivalley excitons and a band of low-energy dark excitons. Such electrical control of optical properties holds promise for various applications in quantum information \cite{henriques2022absorption}, storage, detection \cite{saleem2023theory}, and emission properties in the THz photon energy range. Finally, we have provided a candidate for confinement of tunable neutral excitons that couple strongly to light.

%%%%%%%%%%%%%%%%%%%%%%%%%%%%%%%%%%%%%%%%%%%%%%%%%%%%%%%%%%%%%%%%%%%%%%%%%%%%%%%%%%%%%%%%%%

%%%%%%%%%%%%%%%%%%%%%%%%%%%%%%%%%%%%% Acknowledgments %%%%%%%%%%%%%%%%%%%%%%%%%%%%%%%%%%%%

\section{Acknowledgments}
This research was supported by NSERC Discovery Grant No. RGPIN 2019-05714, the QSP-078 project of the Quantum Sensors Program at the National Research Council of Canada, the HTSN-341 project of the High-Throughput \& Secure Networks Challenge Program at the National Research Council of Canada, University of Ottawa Research Chair in Quantum Theory of Materials, Nanostructures, and Devices., and computing resources at the Digital Research Alliance Canada. GB and YS acknowledge support by the Cluster of Excellence “Advanced Imaging of Matter” of the Deutsche Forschungsgemeinschaft (DFG) - EXC 2056-Project 390715994. K.S acknowledges financial support from National Science Centre, Poland, under Grant No. 2021/43/D/ST3/01989.

%%%%%%%%%%%%%%%%%%%%%%%%%%%%%%%%%%%%%%%%%%%%%%%%%%%%%%%%%%%%%%%%%%%%%%%%%%%%%%%%%%%%%%%%%%

%%%%%%%%%%%%%%%%%%%%%%%%%%%%%%%%%%%%%%%% Contribution %%%%%%%%%%%%%%%%%%%%%%%%%%%%%%%%%%%%

%\section{Author Contributions}

%%%%%%%%%%%%%%%%%%%%%%%%%%%%%%%%%%%%%%%%%%%%%%%%%%%%%%%%%%%%%%%%%%%%%%%%%%%%%%%%%%%%%%%%%%

%%%%%%%%%%%%%%%%%%%%%%%%%%%%%%%%%%%%%%%% Appendix A %%%%%%%%%%%%%%%%%%%%%%%%%%%%%%%%%%%%%%

%%%%%%%%%%%%%%%%%%%%%%%%%%%%%%%%%%%%%%% Bibliography %%%%%%%%%%%%%%%%%%%%%%%%%%%%%%%%%%%%%

\newpage*
\bibliographystyle{apsrev4-2}
\bibliography{Bibliography}

\end{document}